# Large diameter multiwall nanotubes of $MgB_2$: structural aspects and stability of superconducting nanotubular magnesium boride

Pavol Baňacký[1*], Pavol Noga[2], Vojtech Szöcs[1] and Jozef Noga[3,4]

[1]Chemical Physics Division, Institute of Chemistry, Faculty of Natural Sciences, Comenius University, Mlynska dolina CH2, 84215 Bratislava, Slovakia
[2]Advanced Technologies Research Institute, Faculty of Materials Science and Technology, Slovak University of Technology, 91724 Trnava, Slovakia
[3]Department of Inorganic Chemistry, Faculty of Natural Sciences, Comenius University, Mlynska dolina CH2, 84215 Bratislava, Slovakia
[4]Institute of Inorganic Chemistry, Slovak Academy of Sciences, 84536 Bratislava, Slovakia

**Abstract**

Based on a theoretical study, we demonstrated that magnesium boride nanotubes can reach the same stability as bulk $MgB_2$ structure. However, most stable nanotubular forms are not structurally derived from mixed triangular-hexagonal structural motifs of a single layer sheet, which is thought to be the 2D precursor form of all boron nanotubes. $MgB_2$ multiwall nanotubular structures that are derived from multilayer $MgB_2$ slabs with honeycomb B-networks in hexagonal lattices are more stable. The results of an *ab initio* study of multilayer slabs of $MgB_2$ show that a 25-layer slab approaches the stability of bulk $MgB_2$. The critical parameter of the corresponding multiwall nanotubes is the inner diameter; the calculated value is ~ 32.6nm, which is independent of the number of walls. The outer diameters of 25-wall nanotubes are ~ 51 nm, and terminal Mg atoms are located on the outer surfaces of the nanotubes. The electronic band structures of $MgB_2$-multiwall nanotubes ($MgB_2MWNT$) correspond to the band structure character and topology of superconducting bulk $MgB_2$. The results confirm that the quasi-1D superconductor $MgB_2MWNT$ is a stable structure and can be synthesized.

---

[*] corresponding author, banacky@fns.uniba.sk

# 1. Introduction

Control of the chirality of carbon nanotubes (CNT) during synthesis is an unresolved problem, and the production of CNTs with the required electronic properties, i.e., metallic or semiconducting with a medium or small gap [1-3], is unrealistic. Large differences in the conductivities of these forms yields nonuniformity in their emissions and restrict application, especially in the area of field emission. Under these circumstances, a more robust material with a nanotubular form, that does not suffer from these problems, is desired. Boron nanotubes (BNT) or metal borides nanotubes (MeBNT) should not only provide an alternative to CNTs, but also be more efficient. The latest theoretical results [4] indicate that BNTs with a mixed triangular-hexagonal (MTH) structural motif with a diameter > 0.6 nm are thermally stable at temperatures relevant for synthesis, and all of these BNTs are metallic and highly conductive irrespective of their diameters and chiralities [5-9]. These results also agree with the conductivity data for large diameter multiwall BNTs (MWBNT), which have recently been successfully synthesized [10].

The electron deficiency of the B atom gives rise to competition between two and three-centered bonds, resulting in a wide spectrum of hypothetical structures of single-layer (SL) B-sheets with 1-26 atoms/unit cell [4-9, 11, 12-15]. Despite the differences in the underlying unit cells, metal-like band structure (BS) is a common feature of predicted B-sheets. An exception to this rule is the recently predicted semimetallic character [16] of a B-sheet with nonzero thickness (puckered layers in an orthorhombic lattice), which is 50 meV/atom more stable than the $\alpha$-sheet, but remains 320 meV/atom less stable than the bulk rhombohedral $\alpha$-boron structure. However, incorporation of long-range dispersion interactions in Density functional theory (DFT) calculations yields non-puckered BNT conformations with the associated metallic state [17]. These results are compatible with the experimental results from the synthesized MWBNTs, although the morphology of the tubular walls is unknown [10]. Because a layered bulk structure of B does not exist, the synthesis of free-standing SL B-sheets and free-standing single-wall BNTs (SWBNT) is rather problematic. The stability of MWBNT is likely achieved via the contribution of dispersive interactions between neighbouring walls, as shown for DWCNT [18]. Incorporation of metal atoms into BNT can also increase tubular stability. To our knowledge, this effect was for the first time theoretically investigated independently by Quandt et al. [19] (different aspects of BNT formation are summarised in the review paper [20]) and by Chernozatonskii and co-workers [21,22] on small diameter BNTs. Obtained results also indicate possible formation of tubular bundles or double wall nanotubes in dependence on stoichiometry and metal atoms placement into B-network. Stabilization of boron tubular structure by metal atoms incorporation, in particular by Mg atoms, was confirmed experimentally [23,24]. This is important because many metal borides are layered structures and, $MgB_2$ is a high-temperature superconductor with excellent technological parameters. Therefore, knowledge of the structure, morphology, stability and electronic properties of MgBNTs is crucial for the development of superconducting nanoelectronics and nano-optoelectronics.

Here, we show that most stable magnesium boride NTs (MgBNT) are not based on the MTH structural motif of a single layer sheet [11,12], which is supposed to be the 2D precursor form of most stable BNTs [4], but rather are derived from a multilayer $MgB_2$ slab based on the $AlB_2$ structural type with a honeycomb B-network in a hexagonal (hP) lattice. This result is surprising because theoretical studies of a large group of 2D boron sheets with different morphologies predict [12] that the hexagonal pattern of a 2D-boron precursor is one of the most unstable structural forms for Mg incorporation. Based on *ab initio* studies of multilayer slabs of $MgB_2$, we demonstrate that a slab with 25 layers approaches the stability of bulk $MgB_2$. A critical parameter of the resulting $MgB_2$-multiwall nanotube (MgB2MWNT) is the inner diameter, calculated as ~ 32.6nm, which is independent of the number of walls. The outer diameter of a 25-wall NT is ~ 51 nm, and terminal Mg atoms are located on the outer surfaces of the MgB2MWNTs. We show that the electronic band structure (BS) of MgB$_2$MWNT corresponds to the BS character and topology of superconducting bulk MgB$_2$ [25,26].

**2. Calculation methods**

2.1. Electronic structure

Electronic structure calculations on the relaxed geometries of bulk MgB$_2$ and slabs were performed using DFT - band version of the Amsterdam Density Functional (ADF) program package [27] in the LDA and GGA-PBE approximations with a triple-ζ Slater-type basis set with polarization function (TZP) from the ADF basis set library and the Cyclic Cluster HF-SCF (CC HF-SCF) method [28] with the INDO Hamiltonian. The BS topologies of the studied bulk and slab structures obtained by these methods are equivalent. Because the helical symmetry is incorporated only in CC HF-SCF, the BS of the tubular structures, as shown in Figures 1, 3, and 4, were calculated using CC HF-SCF. The dependence of the slab energy on the number of layers with respect to the energy of the bulk calculated using DFT GGA-PBE is equivalent to the dependence calculated using DFT LDA, as shown in Figure 2c.

2.2. Helical symmetry in the band structure calculations of tubular structures

The basic idea of accounting for the helical symmetry in nanotubes, which was suggested in reference [29], more recently in [30, 31] and somewhat differently in [32], was closely adhered to in our implementation. Because some technical details are different and can be written in a simplified manner, we repeated them here to clarify the calculated band structures.

In general, any nanotube with a periodic structure can be constructed by rolling a single sheet of a two-dimensional structure that is finite in one translation direction and infinite in the other one. We shall restrict ourselves to nanotubes created from two-dimensional hexagonal lattices characterized by two equivalent primitive translational vectors $\vec{A}_i$, that contain an angle of $2\pi/3$. For general 2D lattices, treatment is similar with primitive translational vectors of $\vec{A}_1$ and $\vec{A}_2$ containing an angle of α.

Because of our chosen convention, the dimension in the direction of $\vec{A}_1$ will be treated as finite, whereas an infinite number of translations is assumed along $\vec{A}_2$. A nanotube characterized by a general chiral vector (*m, n*) is then created from the sheet that has *m* translations (0, ..., *m* − 1) along $\vec{A}_1$ and "infinite" number of translations ($n_{tr}$) along $\vec{A}_2$. This ribbon is rolled on a cylinder with the diameter $d_{NT} = |m\vec{A}_1 + n\vec{A}_2|/\pi$, which follows from the fact that the chiral vector is rolled perpendicular to the rotation axis, forming the circumference of the cylinder. The irreducible computational unit cell corresponds to the one in the two-dimensional structure, except for the geometry relaxation because of the curvature. Similar to the two-dimensional structure, which is infinite in two dimensions, each unit cell experiences the same environment. Original translations along $\vec{A}_i$ are now transformed to roto-translations ($\hat{\tau}_i$), which are characterized by the pair of operations ($\vec{a}_i, \varphi_i$), where $\vec{a}_i$ is the projection of $\vec{A}_i$ onto the axis of the nanotube and $\varphi_i$ is the rotation angle related to this translation. Therefore, for any point defined in a cylindrical coordinate system (ρ, φ, z), $\hat{\tau}_i \equiv (\rho, \phi + \varphi_i, z + a_i)$.

If we relate a pseudovector $t_i$ to these roto-translations, analogous to the two-dimensional planar lattice, we can define a reciprocal pseudovector, $\bar{t}_i$, such that $\bar{t}_i t_j = 2\pi\delta_{ij}$.

The atomic orbital $\chi_{j_1, j_2}$ is a counterpart of the reference unit cell atomic orbital $\chi_{0,0}$ in the unit cell defined by $\hat{\tau}_1^{j_1}$ and $\hat{\tau}_2^{j_2}$ roto-translations. The structure created by *m* roto-translations $\hat{\tau}_1$ (including 0) of the reference computational cell can be treated as an ideal cyclic cluster with periodic boundary conditions, i.e., each nanotube unit has an equivalent surrounding. Consequently, from *m* atomic orbitals $\chi_{j_1, 0}$ ($j_1$ = 0, *m* − 1) one can create *m* symmetry orbitals:

$$\chi^{\kappa_l} = \frac{1}{\sqrt{m}} \sum_{j_1=0}^{m-1} e^{i\kappa_l R_{j_1}} \chi_{j_1, 0} \quad ,$$

where $R_{j_1} = j_1 t_1$ and there are *m* allowed discrete values of $\kappa_l = \frac{l}{m}\bar{t}_1$ for $l = 0, m-1$. These symmetry orbitals are propagated because of the $\hat{\tau}_2$ operation to infinity, providing Bloch

orbitals:

$$\chi^{(k,\kappa_l)} = \lim_{N\to\infty} \frac{1}{\sqrt{m.N}} \sum_{j_2=-N/2}^{N/2} e^{i(k.R_{j_2}+\kappa_l R_{j_1})} \chi_{j_1,j_2}$$

where $R_{j_2} = j_2 t_2$ and $k$ is any value from the first Brillouin zone for the one-dimensional system. In the practical implementation into the routines that generate integrals in Cartesian coordinate systems, one has to consider the appropriate rotation of the coordinate system for each basis function of $p, d, f...$ type to preserve the rotational symmetry.

Because the present work only reports nanotubes with chiral vectors $(m,0)$, we denote $\kappa$ as $\kappa_{\phi_r}$ in the main text because it is related the true rotation angle in the $m$-fold rotational symmetry, whereas $k$ is denoted $k_{tr}$ and $m = n_a$. In this special case the rotational symmetry can also be accounted for through the point symmetry operations [32].

## 2. Results and Discussion

As predicted previously [12], the most stable 2D precursor for MgBNT formation is an SL sheet of MTH character (hole density $\eta=1/4$) with a unit cell composition of $B_{18}Mg_9$, shown in **Figure 1a**. The study of single-wall MgBNT (SWMgBNT) based on this SL complex structure, has not yet been published. Here, we present basic results for the first time. Accordingly, because of the interplay between core repulsion and one and two-electron contributions of inner and outer Mg atoms, the folding energy has a nontrivial dependence on the NT diameter. NTs (**Figure 1b**) with diameters in the range of 2 to 4 nm, are most stable, with extreme at 2.3 nm, as shown in **Figure 1c.** The electronic BS of these NTs are metallic, as shown in **Figure 1d**. Despite the fact that Mg stabilizes SWMgBNT formation (~ -0.25 eV / $MgB_2$), the parent $B_{18}Mg_9$ sheet is ~1.3 eV/ $MgB_2$ less stable than bulk $MgB_2$.

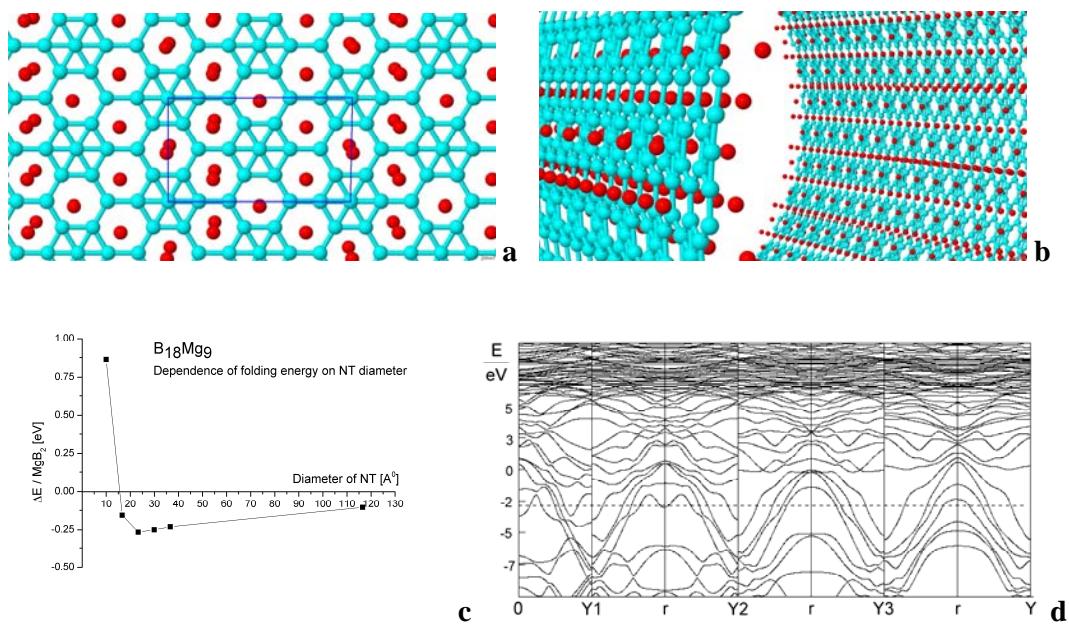

**Figure 1.** (colour online) **Structure of $B_{18}Mg_9$** (**a**) 2D sheet indicating the rectangular unit cell and (**b**) single-wall $B_{18}Mg_9$ nanotube constructed via a helical method. Six Mg atoms (red) are placed above and below the centres of 3 empty hexagons, and the remaining 3 atoms are located above the centres of the next 3 hexagons. The dependence of nanotube stability on diameter is shown in panel (**c**) – reference 0 eV is the energy of the 2D sheet. The electronic band structure of the most stable nanotube, which has a diameter of 2.33 nm ($n_a$=7) for allowed rotation wave numbers $k_{\Phi r}=r/n_a$ with $r = 0$, $r = \pm 1$, $\pm 2$, $\pm 3$ in paths arrangement [($-r/n_a$,1/2)→($-r/n_a$,0) | ($r/n_a$,0) →($r/n_a$,1/2)], is shown in panel (**d**). The dotted line is the Fermi energy.

Reports on the synthesis of MgBNT are very rare [23, 24, 33, 34]. Although many sophisticated experimental methods have been developed, it is remarkable that a simple thermal evaporation of $MgB_2$ powder under a special rapid heating protocol without a catalyst can produce nanoparticles (NP) in an 80% yield [24]. Applied structural methods revealed that the individual NPs are not nanowires but MWNTs with an inner diameter of ~30 nm, outer diameter of ~90 nm and a length of several microns. Magnetic susceptibility measurements have shown that the as prepared MWNTs are superconductors with better parameters than the starting powder. The stoichiometric composition of Mg:B=1:2, was indicated by elemental microanalysis. However, the atomic structure and morphology of individual walls in the synthesized MgBMWNT remained unknown. Thus, the question arises whether the existence of a superconducting nanotubular form of $MgB_2$ is possible or whether the prepared MgBMWNTs have some other structure.

The Mg:B=1:2 composition, i.e., $(MgB_2)_x$-MWNT, is a challenge for theory related to the structural specification of hypothetical SL precursor, and mainly for studying the stability, structure and electronic properties of the corresponding MWNT. Formally, the most stable SL precursor predicted [12], $B_{18}Mg_9$, matches the required stoichiometry - $(MgB_2)_9$. However, because Mg-atoms are placed on both sides of the B-plane, $B_{18}Mg_9$ is basically excluded as a layered 2D precursor, giving rise to a stable MW structure. For the same reason, other proposed structures [12] with MTH character; hole densities of $\eta$=1/13, 1/9, 1/7, 1/5, and 3/10; and stoichiometries of $(MgB_2)_x$, are disqualified as 2D precursors for MWNT formation.

Thus, hexagonal sheet ($\eta$=1/3) in an hP lattice derived from bulk $MgB_2$, as shown in **Figure 2a**, is the only option. This SL structure is ~0.8 eV/$MgB_2$ less stable than the $B_{18}Mg_9$ sheet and more than 2.1 eV/$MgB_2$ less stable than bulk $MgB_2$. Nonetheless, in contrast to the sheets that were predicted to be much more stable, this structure allows for the formation of stable slabs in the (001) direction. In **Figure 2b**, a 5-layer slab and a slice of a 5WNT created from such slab are shown in **Figure 2d**.

We studied the dependence of slab stability on the number of layers, as shown in **Figure 2c**. Both applied methods (DFT [27] and CC HF-SCF [28] – Section 2.1.) show that a slab with 25 layers approaches the stability of bulk $MgB_2$. As the numbers of layers increases, the stability of the slab approaches the bulk limit and the electronic BS of the slab evolves toward the BS topology of bulk $MgB_2$.

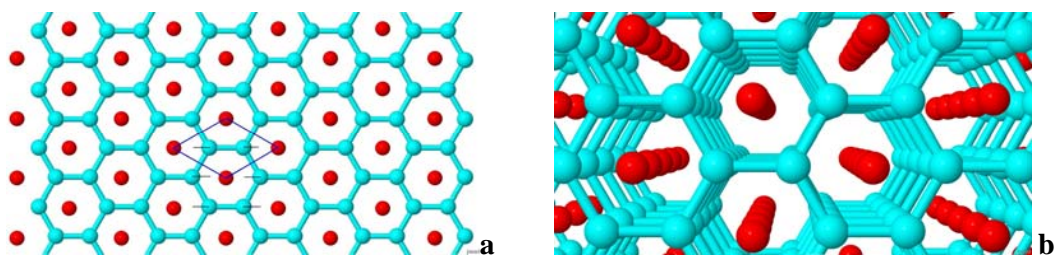

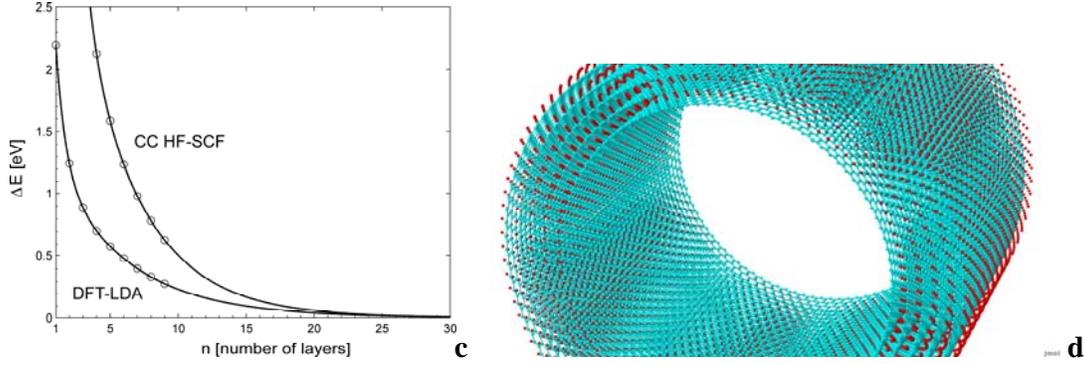

**Figure 2.** (colour online) **Structure of MgB$_2$;** (**a**) 2D sheet indicated the hP unit cell and B-B stretching in E$_{2g}$ mode and (**b**) 5-layer slab in the (001) direction. The dependence of the slab electronic energy on the number of layers is shown in panel (**c**) with reference 0 eV, which corresponds to the energy of bulk MgB$_2$. Panel (**d**) depicts a 5-wall MgB$_2$ nanotube constructed from the slab using helical method with terminal Mg atoms on the outer tubular wall.

This trend is shown for the BS of the 5-layer slab in **Figure 3c**; the tops of the degenerated σ-bands represent an increasing number of layers that were increasingly squeezed and shifted toward the Fermi level (FL) with a characteristic hole pocket in the σ-bands that arises at the Γ point. The electronic BS of bulk MgB$_2$ is unstable in electron-phonon (EP) coupling to B-B stretching vibrations in the E$_{2g}$ phonon mode, e.g. see references [35, 36].

For the equilibrium hP-structure of MgB$_2$, where $a_{eq}=b_{eq}$=3.0823Å, $c_{eq}$=3.5146Å and fractional coordinates B1(1/3,2/3,0), B2(2/3,1/3,0), and Mg(0,0,-1/2), we calculated that the σ-band splitting, which is related to fluctuations of one of the σ-bands across FL, occurs at a B-displacements $\Delta d^{cr}/2 = \pm 0.032$Å /atom out of equilibrium in the direction of the B-B stretching mode. This result also holds for bulk as well as for MgB$_2$ slabs (c.f., **Figure 3a-b, 3c-d**). The change in the bond length for this value is slightly smaller than the zero point energy (ZPE) amplitude, $\Delta d^{ZP} \sim 0.068$Å, of the B-B stretching mode in MgB$_2$ [35]. EP coupling to this mode induces small Fermi energy effects, in particular, the transition into a superconducting anti-adiabatic state [35, 36].

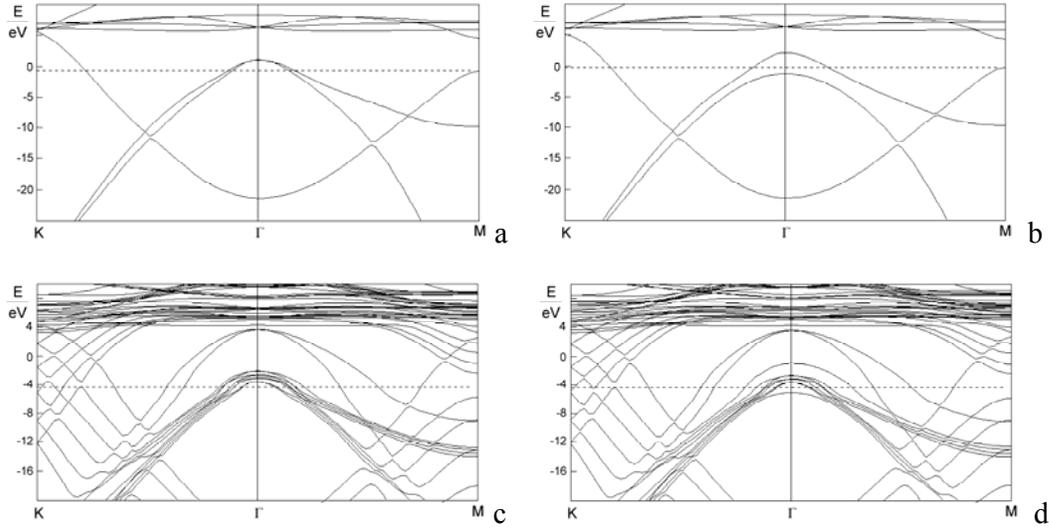

**Figure 3. Electronic band structure of MgB$_2$ (a)** in the equilibrium geometry for the bulk form and **(b)** induced instability, fluctuation of the σ-band across the Fermi level at Γ point with electron-phonon coupling to the E$_{2g}$ phonon mode for B-atom displacements of ±0.032Å /atom in the stretching vibration mode. The **band structure of a 5-layer slab** in equilibrium is shown in panel **(c)** and the band structure for B-atom displacements of ±0.032Å /atom from equilibrium is shown in panel **(d)**.

Therefore, a stable slab can serve as a precursor for MWNT formation and $\Delta d^{cr}$ ($\Delta d^{ZP}$) can be used for the calculation of the critical diameter, $D^{i,cr}$, of superconducting MgB$_2$MWNT. However, in MWNT modelling, a nontrivial problem arises because the original unit cell, which is common in multilayer slabs, no longer exists. During folding, each layer of the parent slab forms a tube with a different diameter; consequently, there is no finite common unit cell for the resulting MWNT. Under these circumstances, instead of a chiral method of NT formation [3], it is more suitable to use a method based on helical symmetry [30], which is implemented into CC HF-SCF code in a modified form [37], see Section 2.2. In the roto-translation helical treatment, the unit cell of the tube is the same as the unit cell of the parent 2D structure, with the same number of bands, and direct correspondence between planar and tubular BS is reached. A tube [($n_a$,0), $n_{tr}$] is characterized by a diameter D = $a \cdot n_a/\pi$, and helical unit angle, $\Phi = 2\pi/n_a$, where $a$ is the lattice parameter of the planar unit cell in the **a** direction ($\boldsymbol{a} \equiv \vec{A}_1$). The multiplicity (fold-number) of the rotation axes symmetry for the created tube is $n_a$ ($n_a \equiv m$). This method can also be applied to a slab, separately for successive layers with increasing (decreasing) $n_a$-values, which gives rise to MWNT formation of concentric tubes with different diameters and different multiplicities for the rotation axes symmetry but with the same helical tubular axes (e.g., Figure 2d). Although applicable in principle, real calculation of a 25-wall NT would not only be extremely time and memory demanding but convergence of the electronic energy would be problematic. Instead, we can indirectly calculate the basic properties of MgB$_2$MWNT.

At equilibrium, the B-B bond length, $d_{BBeq}$, is 1.7795 Å, whereas due to ZPE vibrations, $d_{BB} = d_{BBeq} \pm \Delta d^{ZP}$. The threshold values of $d_{BB}$ correspond to some hypothetical lattice constants, $a = a_{eq} \pm \Delta a$, where $\Delta a = \Delta d \sqrt{3}$. To preserve superconductivity, the character of the σ-bands and BS topology of the created MgB$_2$MWNT have to be as close as possible to those of the bulk material. Because of the largest curvature, the diameter of the inner wall is critical because the BS topology of this wall is most prone to deformation. Therefore, we require no deformed BS in configuration space $a_{eq} \pm \Delta a^{cr}$, or at least $\pm \Delta a^{ZP}$. Then, for the tubular inner wall and two successive walls, the following coupled sets of equations should hold;

A/ $(a_{eq} - \Delta a/2).n_a^i = 2\pi.r^i$, $a_{eq}.(n_a^i + 1) = 2\pi(r^i + c_{eq})$ for the inner (1$^{st}$) and 2$^{nd}$ walls, which has a solution of $n_a^i = (2\pi.c_{eq} - a_{eq})/(\Delta a/2)$, and

B/ $a_{eq}.(n_a^i + 1) = 2\pi(r^i + c_{eq})$, $(a_{eq} + \Delta a/2).(n_a^i + 2) = 2\pi(r^i + 2c_{eq})$ for the 2$^{nd}$ and 3$^{rd}$ walls, which has a solution of $n_a^i = (2\pi.c_{eq} - (a_{eq} + \Delta a))/(\Delta a/2)$.

The coupled sets are exact only within the limit of $\Delta a \rightarrow 0$, i.e., for planar slab ($n_a^i \rightarrow \infty$, $\Phi \rightarrow 0$). The curvature of the walls is the reason that $\Delta a \neq 0$, and the requirements (A, B) for 3 successive inner walls restricts change in $a_{eq}$ by no more than $\pm \Delta a/2$. An approximate solution for critical $n_a^{i,cr}$, calculated as an integer mean value for $\Delta a^{cr} = \Delta d^{cr} \sqrt{3}$ and $\Delta a^{ZP} = \Delta d^{ZP} \sqrt{3}$, is $n_a^{i,cr} = 332$. This value represents the multiplicity of the rotation axes symmetry of the inner wall with a critical-minimal inner diameter for the MWNT, $D^{i,cr} = n_a^{i,cr} \cdot a_{eq}/\pi \sim 32.6$ nm, and a helical unit angle, $\Phi^{cr} = 360/n_a^{i,cr} \sim 1.08434°$. Toward the outer surface, the curvature of the walls decreases, which is related to the increase in the multiplicity of rotation axes symmetry of a particular wall. For inner diameter $D^{i,cr} \sim 32.6$ nm, the 25-layer slab yields an outer NT diameter of $D^{out} = D^{i,cr} + 50 \cdot c_{eq} \sim 51$ nm. With increasing diameter, the folding energy decreases and the energy of the NT approaches the energy of parent structure. Then, it is reasonable to expect that the energy of MgB$_2$25WNT with $D^{i,cr} \geq$ ~32.6 nm approaches the stability of a 25-layer slab i.e., bulk MgB2, as shown in Figure 2 c. More over, the curvature of the most important inner wall is small enough in order to preserve the BS character and topology, which is crucial for transition into a superconducting state. As shown in **Figure 4**, using the calculated $D^{i,cr}$, the NTs exhibit the correct BS with only a few walls, although they are substantially less stable than the 25-WNT.

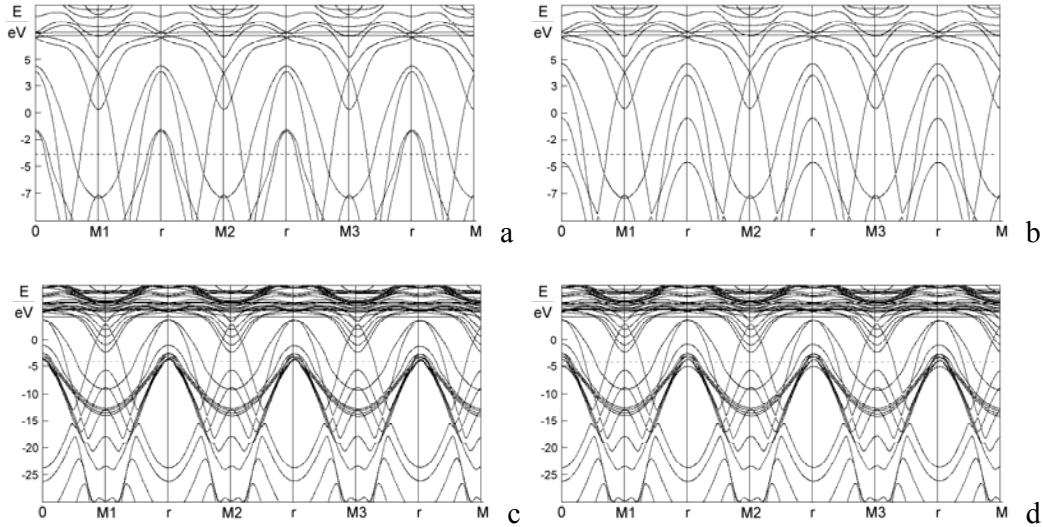

**Figure 4. Electronic band structure** of a 2-wall MgB$_2$ nanotube (**a**) at equilibrium geometry and (**b**) with an induced instability, fluctuation of the σ-band across the Fermi level with electron-phonon coupling to the E$_{2g}$ phonon mode for B-atom displacements of ±0.032Å /atom in the stretching vibration mode. The band structure of a 5-wall MgB$_2$ nanotube at equilibrium is shown in panel (**c**) and the band structure with B-atom displacements of ±0.032Å /atom from equilibrium is shown in panel (**d**). The electronic band structures are for nanotubes with critical diameters of 32.6 nm (n$_a$= 332) for the fewest allowed rotation wave numbers, i.e., $k_{\Phi r}=r/n_a$ with $r = 0$, $r = \pm1, \pm2, \pm3$ in the paths arrangement $[(-r/n_a,1/2)\rightarrow(-r/n_a,0) | (r/n_a,0) \rightarrow(r/n_a,1/2)]$.

The single-wall MgB$_2$NT created from SL MgB$_2$ is the most energetically unfavourable, as shown in Figure 2c. Interestingly, even the least stable, MgB$_2$SWNT calculated independently using the Generalized Force Field method and a single-layer fragmental model of MgB$_2$, also yields a large diameter NT with an energy minimum for the tubular diameter ~ 30.6 nm [33].

With regard to Mg atom placement, our calculations showed that MWNTs are more stable when Mg atoms are on the outer tubular surface. More over, this configuration is more stable than the corresponding slab. In the case of MgB$_2$2WNT with $D^{i,cr}$ = 32.6 nm, it is ~ - 0.16 eV/MgB$_2$ compared to the 2-layer slab. This result suggests that MgB$_2$MWNTs can grow on the surface of a proper substrate by first capturing Mg atoms with energies less than the binding energy of Mg to the B-network. The first layer can grow in the (001) direction, forming a slab that can begin to curl if the unit cells parameters of the substrate and MgB$_2$ do not match sufficiently, where the first layer becomes the outer wall of the emergent MWNT. The formation of MgB$_2$MWNTs should be kinetically driven and sensitive to not only the rate of heating of the MgB$_2$ powder but also the subsequent cooling regime and character of the substrate. In particular, the binding energy of Mg to surface of hexagonal BN was calculated to be -0.057 eV. This result indicates that the rate of cooling at T~ 300° - 100° C should be critical if hexagonal BN is used as the substrate.

## 3. Conclusion

Our results suggest that the quasi-1D superconductor, MgB$_2$MWNT, is a stable structure and can be synthesized, as experimentally indicated by Zhou et al. [24]. This material should be excellent for application in nanoelectronics. The single-photon detectors, based on this material, could be expected with fast optical response at relatively high operating temperature, and should become important elements in the development of quantum computing technologies [38].

**Acknowledgements**

This work was supported by the Slovak Research and Development Agency under contract No. APVV-0201-11. The authors thank Uniqstech a.s. for the use of the Solid2000-NT source code for this research.

**References**

1. T.W. Odom, J-L. Huang, P. Kim and, Ch.M. Lieber, *Nature* **1998**, 391, 62.

2. J.W.G. Wilder, L.C. Venema, A.G. Rinzler, R.E. Smalley and, C. Dekker, *Nature* **1998**, 391, 59.

3. R. Saito, G. Dresselhaus and, M.S. Dresselhaus, *Physical properties of carbon nanotubes,* ICP, London, **1998**.

4. J. Kunstmann, V. Bezugly, H. Rabbel, M.H. Rümmeli and, G. Cuniberti, *Adv. Funct. Mater.* **2014**, 24, 4127.

5. I. Boustani, A. Quandt, E. Hernandez and, A. Rubio, *J. Chem. Phys*. **1999**, 110, 3176.

6. J. Kunstmann and, A. Quandt, *Phys. Rev. B* **2006**, 74, 035413.

7. C.K. Lau, R. Pati, R. Pandey, A.C. Pineda, *Chem. Phys. Lett.* **2006**, 418, 54.

8. V. Bezugly, H. Eckert, J. Kunstmann, F. Kemmerlic, H. Meskine, G. Cuniberti, *Phys. Rev. B* **2013**, 87, 245409.

9. V. Bezugly, J. Kunstmann, B. Grundkotter-Stock, T. Frauenheim, T. Niehaus, G. Cuniberti, *ACS Nano* **2011**, 5, 4997.

10. F. Liu, C. Shen, Z. Su, X. Ding, S. Deng, J. Chen, N. Xu and, H. Gao, *J. Mater. Chem.* **2010**, 20, 2197.

11. H. Tang, S. Ismail-Beigi, *Phys. Rev. Lett.* **2007**, 99**,** 115501.

12. H. Tang, S. Ismail-Beigi, *Phys. Rev. B*. **2009**, 80, 134113.

13. X. Wu, J. Dai, Yu. Zhao, Zh. Zhuo, J. Yang and, X.Ch. Zeng, *ACS NANO* **2012**, 6**,** 7443.

14. X. Yu, L. Li, X-W. Xu and, Ch.Ch. Tang, *J. Phys. Chem. C* **2012**, 116**,** 20075.

15. Z.A. Piazza, H-S. Hu, W.L. Li, Y-F. Zhao, J. Li and, L-Sh. Wang, *NATURE COMMUNICATIONS* **2014**,| 5:3113 | DOI: 10.1038/ncomms4113.

16. X-F. Zhou, X. Dong, A. R. Oganov, Q. Zhu, Y. Tian, H-T. Wang, *Phys. Rev. Lett.* **2014**, 112**,** 085502.

17. R.N.Gunasinghe, C.B. Kah, K.D. Quarles, X.Q. Wang, *Appl. Phys. Lett.* **2011**, 98, 261906.

18. K. Liu, Ch. Jin, X. Hong, J. Kim, A. Zettl, E. Wang and, F. Wang, *NATURE PHYSICS* **2014**, 10, 737.

19. A. Quandt, A.Y. Liu, I. Boustani, *Phys. Rev. B.* **2001**, 64**,** 125422.

20. A. Quandt, I. Boustani, ChemPhysChem. **2005**, 6, 2001.

21. L.A. Chernozatonskii, *JETP Letters* **2001**,74, 335.

22. P.B. Sorokin, L.A. Chernozatonskii,| P.V. Avramov , B. I. Yakobson, *J. Phys. Chem. C***. 2010,** 114, 4852.

23. E. Iyyamperumal, F. Fang, A.B. Posadas, Ch. Ahu, R.F. Klie, Y. Zhao, G.L. Haller, L.D.Pfefferle, *J. Phys. Chem. C* **2009**, 113,17661.

24. Sh-M. Zhou, P. Wang, Sh. Li, B. Zhang, H-Ch. Gong, X-T. Zhang, *Materials Letters* **2009**, 63, 1680.

25. J. Nagamatsu, N. Nakagawa, T. Muranaka, Y. Zenitani and, J. Akimitsu, *Nature* **2001**, 410, 63.

26. H.J. Choi, D. Roundy, H. Sun, M.L. Cohen, S.G. Louie, *Phys. Rev. B* **2002**, 66, 020513.

27. Computer code, *ADF2014.99, SCM*, Theoretical Chemistry, Vrije Universiteit, Amsterdam, The Netherlands

28. J. Noga, P. Banacky, S. Biskupic, R. Boca, P. Pelikan, M. Svrcek, A. Zajac, *J. Comp. Chem*. **1999**, 20, 253.

29. C.T. White, D.H. Robertson, J.W. Mintmire, *Phys. Rev. B* **1993**, 47, 5485.

30. P.N. D'yachkov, D.V. Makaev, *Phys. Rev. B* **2007**, 76, 195411.

31. C.T. White, J.W. Mintmire, *J. Phys. Chem. B* **2005,** 109, 52.

32. Y. Noel, P. D'Arco, R. Demichelis, C.M. Zicovich-Wilson, R. Dovesi, *J. Comp. Chem.* **2010**, 31, 855.

33. N. Sano, O. Kawanami, H. Tamon, *J. Appl. Phys*. **2011**, 109, 034302.

34. F. Fang, E. Iyyamperumal, M.F. Chi, G. Keskar, M. Majewska, F. Ren, C.C. Liu, G. L. Haller, L.D. Pfefferle, *J. Mat. Chem. C* **2013**, 1, 2560.

35. L. Boeri, E. Capellutti, L. Pietronero, *Phys. Rev. B* **2005**, 71, 012501.

36. P. Banacky, *J. Phys. Chem. Solids* **2008**, 69, 2696.

37. P. Banacky, J, Noga, V. Szocs, *J. Phys. Chem. Solids* **2012**, 73**,** 1044.

38. Ch.M. Natarajan, M.G. Tanner, R.H. Hadfield, *Supercond. Sci. Technol.* **2012**, 25, 063001.